# Architecture Proposal for 6G Systems Integrating Sensing and Communication

Peter Gersing[1)], Mark Doll[2)], Joerg Huschke[3)], Jan Plachy[4)], Stefan Köpsell[5)]

[1)]GPP Communication, Oberhaching, Germany

[2)]Nokia Bell Labs, Stuttgart, Germany

[3)]Ericsson, Herzogenrath, Germany

[4)]Deutsche Telekom AG, Berlin, Germany

[5)]Barkhausen Institut, Germany

**Abstract**

Integrating sensing functionality into 6G communication networks requires some changes to existing components as well as new entities processing the radar sensing signals received by the communication antennas. This whitepaper provides a comprehensive overview of the 6G design proposal for ISaC (Integrated Sensing and Communication). The whitepaper has been created by the architecture group of the KOMSENS-6G project with the intend to serve as a basis for further discussions and alignment across innovative 6G projects.

## Introduction

The KOMSENS-6G project develops a potential 6G ISaC architecture, along with use cases, protocols and the processing of sensing data. The ISaC service is already in discussions in 3GPP, a feasibility studies identified a set of use cases, derived first requirements and a studies on stage 2 as well as security and privacy requirements are ongoing [1,2,3,4]. The existing 3GPP architecture is taken as a basis to propose the 6G architectural concepts for integrating "Sensing" services in communication networks.

This whitepaper addresses the ISaC architecture mainly on logical level. Some implementation and deployment aspects are being included as far as they are relevant to further discussions.

The KOMSENS-6G project is focusing on network-based sensing, whereas sensing via the UE (User Equipment) is out of scope.

## The KOMSENS-6G SysML Model

The architecture group has developed a SysML[1] model (System Design and Modelling Language) of the envisioned ISaC architecture for 6G networks. SysML allows the definition of the system structure as well as the system behavior, including interfaces, message-sequences, data formats and more. Modeling the system supports the consistency of the architecture concepts by serving as a single point of truth. The diagrams of this whitepaper are generated from the KOMSENS-6G SysML Model.

For easier orientation in the terminology, high level description of ISaC notations, based on ongoing 3GPP studies [3] is provided in Table 1.

---

[1] The SysML language is a standardized graphical language to support the specification of the system structure and the system behavior. The standard is maintained by the OMG (Object Management Group) – https://www.omgsysml.org/.

## Logical System Architecture for ISaC

Receiving and processing reflected signals (sensing) by communication antennas is the new sensing feature which is to be integrated into the 6G communication systems. The basis for the logical architecture shown in Figure 1 is the current 3GPP 5G system architecture [5], to which we propose modifications to integrate sensing (ISaC). The integration can be accomplished by enhancing selected existing components (blue) and adding new ISaC components (red). Further architectural options deviating from the baseline have been discussed in the project recently and are being addressed alongside the discussion of the baseline that is following hereafter.

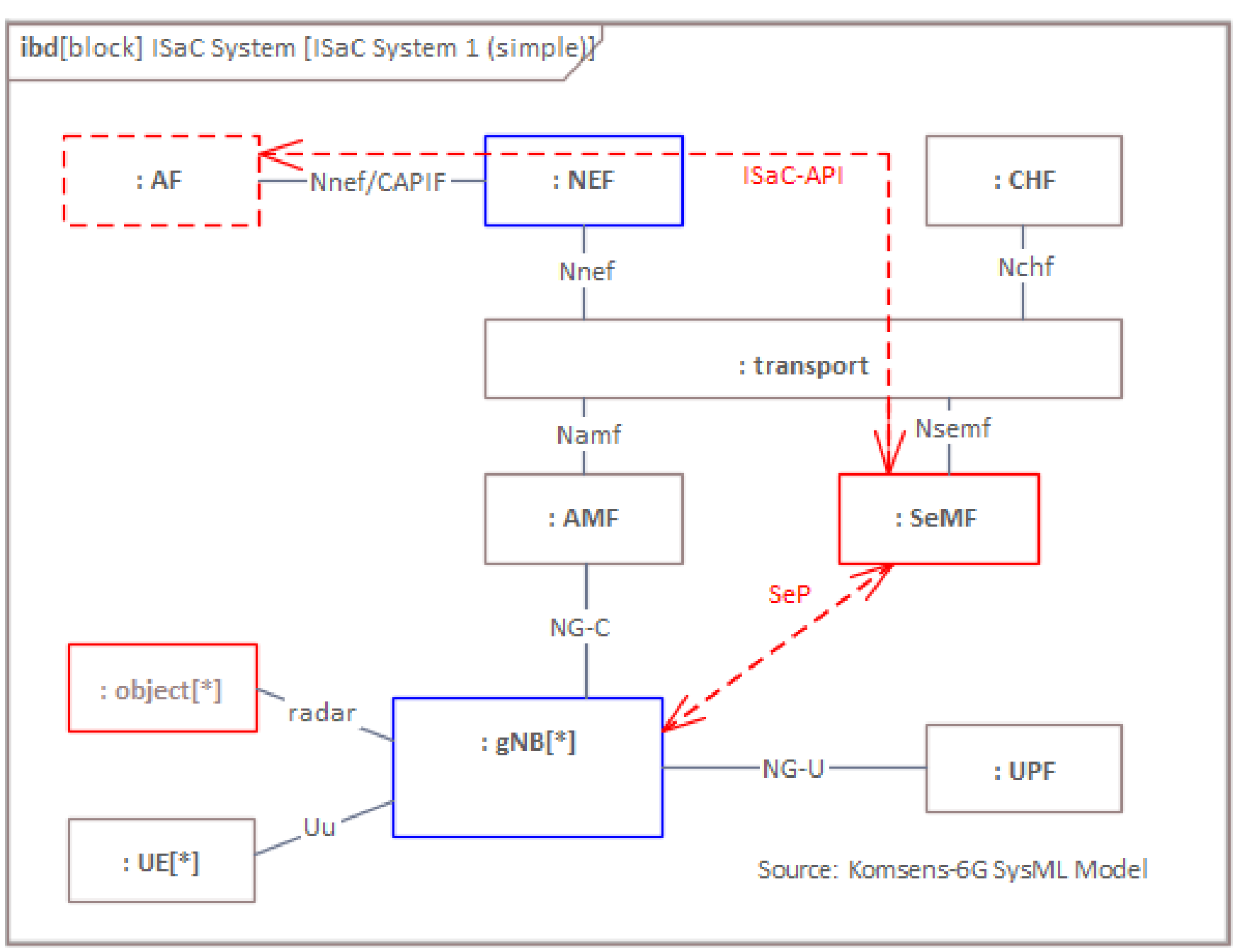

*Figure 1: Logical Architecture for ISaC.*

The RAN's base stations (gNB, next Generation NodeB) are enabled to process sensing signals. The new SeMF (Sensing Management Function) in the core implements sensing tasks that are preferably centralized and manages the overall sensing operation. Furthermore, SeMF communicates with CHF (Charging Function) for sensing service charging. Note that charging is expected to be based on the requirements (Sensing Quality of Service (SQoS), coverage, periodicity, duration, etc.) of the sensing request, as these determine required resource allocation and processing, while sensing result transport may follow common data transport charging.

*Table 1. ISaC notations.*

| ISaC notations | Description |
|---|---|
| **Sensing Entity** | The Sensing Entity referring to a Sensing Transmitter and/or to a Sensing Receiver. Note we assume only gNB as Sensing Entity in KOMSENS-6G. |
| **Sensing Function/Sensing Management Function (SeMF)** | Indicating the logical function which is involved to support Sensing Service. |
| **Sensing Service** | Capability to collect and provide information about object and/or characteristics of the environment using radio signals. |
| **Sensing data** | Data derived from radio signals impacted by an object or environment of interest for sensing purposes, and optionally processed within the gNB before being transported to SeMF. |
| **Sensing result** | Processed sensing data requested by a service consumer and exposed via NEF. |
| **Sensing Service Consumer** | The entity that may consume the Sensing Result. The Sensing Service Consumer may also request the Sensing Result. |
| **Sensing contextual information** | Information that is exposed with the sensing results by 5G system to a trusted third-party which provides context to the conditions under which the sensing results were derived. |

The sensing service will be provided to users via a new ISaC-API (Application Programming Interface). The new protocols for sensing as shown in Figure 1 as red dashed lines with arrows on both ends are:

- the ISaC-API for exposing the sensing service and providing sensing results through an application function (AF) to Sensing Service Consumer, and
- the SeP (Sensing Protocol) for the communication (sensing control and sensing data) between the core (SeMF) and the RAN (gNB).

To support ISaC-API the NEF (Network Exposure Function), which interfaces the core with applications like a Sensing AF (Application Function) using the sensing service, is extended.

The logical communication between the core and the radio is defined by the protocol SeP (as specified in this Whitepaper). This includes the messages for the control of the sensing entities in the RAN as well as providing sensing data from gNB(s) to the requesting SeMF. Multiple options are being discussed for the transport of SeP between gNB and SeMF. The options include re-using existing interfaces and protocols: sensing control to be transported over AMF, leveraging NRPPa (New Radio Positioning Protocol Annex) or a similar transport for SeP, and sensing data either the same as control via AMF, but which could impact latency of other control, or UPF (User Plane Function). KOMSENS-6G focuses on 6G, and thus, with more flexibility new options (ordered by the level of change needed and increasing extendibility) for the transport of sensing control and sensing data between sensing entity (gNB) and SeMF are in discussion (details in Table 2):

1. Sensing control and data transported via a new point-to-point interface between gNB and SeMF. This option is currently being standardized in 3GPP for 5G Advanced [6].
2. Sensing control through AMF (in control plane) and sensing data transported in data plane (being discussed for AI/ML at this time) that can be shared for transport of sensing data, non-control and non-user data.
3. Sensing control and sensing data transported via SBI (Service-Based Interface) from gNB to SeMF), to follow the SBA (Service-Based Architecture) of the 5G core network.

*Table 2: KOMSENS-6G considered options for transport of sensing data and sensing control.*

| | **Option 1** | **Option 2** | **Option 3** |
|---|---|---|---|
| **Sensing control** | New P2P interface | AMF | SBI to Core |
| **Sensing data** | New P2P interface | Data plane interface | SBI to Core |
| **Required modifications** | New interface between RAN and Core | New interface between RAN and Core | Transition to RAN ↔ Core SBA |
| **Benefits** | Separation of Sensing | Reuse of existing CP & Unified data transport | Flexibility of RAN ↔ Core |
| **Differentiation of sensing data from other user plane traffic** | Yes | Yes | Yes |
| **Drawbacks** | New (dedicated) interface | New (shared) data plane | Paradigm shift |
| **Extendibility** | limited | good | highest |
| **Required architecture changes** | minimal | limited | extensive |

Note that we do not discuss to integrate sensing into 5G. We use the 5G terminology here solely as placeholders for future 6G architecture functional entities, as those have not yet been defined.

**Sensing Modes**

There are two major variants of sensing: monostatic sensing where the sensing TX and RX are co-located, and bistatic sensing where the TX and RX are on different sites. In monostatic sensing, the self-interference from the TX spilling over into the RX is a serious challenge that requires additional hardware efforts in terms of TX-RX isolation or self-interference cancellation. Accordingly, instead of a strictly monostatic setup with a single Transmission-Reception Point (TRP) acting as both TX and RX, a quasi-monostatic setup with two separate TRPs for TX and RX closely co-located at the same site can be used. In bistatic sensing the interference level at the RX can be controlled by the separation to the TX and is manageable without major hardware efforts.

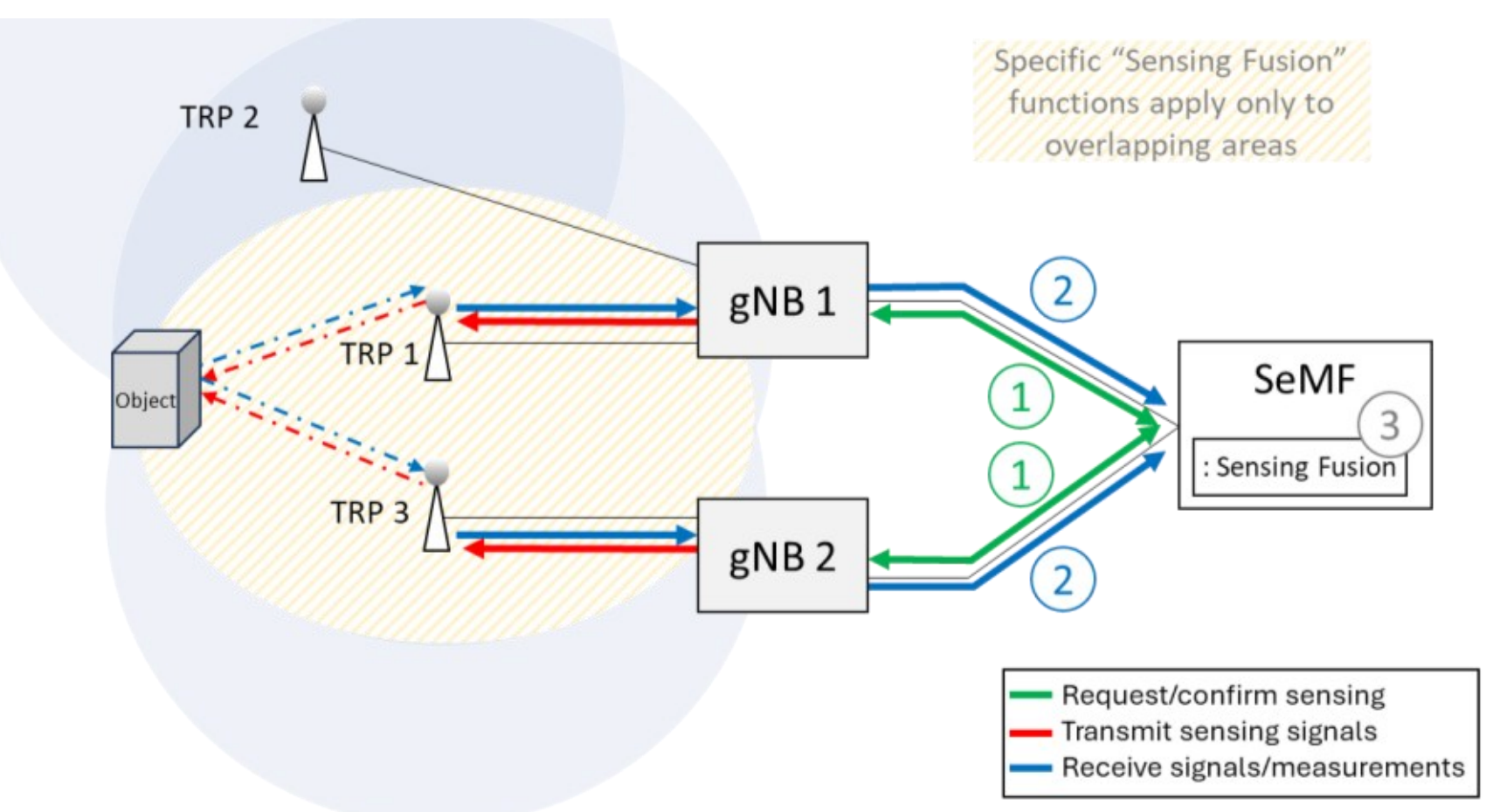

*Figure 2: Multi-monostatic sensing.*

For multi-monostatic sensing multiple TRPs are being used, measuring in monostatic mode concurrently in an overlapping area as illustrated in Figure 2. In multi-monostatic sensing, each gNB is individually configured by the SeMF through parallel independent signaling procedures. First, the SeMF requests sensing measurements from selected TRPs by a sensing request to the respective gNB. Then accordingly each TRP transmits its own distinct sensing signal. Details on scheduling are found in section Control Plane further below. From the received signal (back-scattered sensing signal, i.e., the dash dot line with arrows), each gNB computes the sensing data and reports them back to the SeMF. Depending on the TRP's capabilities and the configuration through the SeMF, the gNB may process sensing data to detect back-scatterers and provide their location in 2D or 3D space, or just their distance from the TRP, with or without speed (Doppler) information. The SeMF collects all sensing data, and its "Sensing Fusion" function combines them into a more accurate sensing result of the whole area, respectively in case of range only measurements trilaterates to determine object positions.

A generalization of bistatic sensing is multistatic (or multi-bistatic) sensing, illustrated in Figure 3, where there is one sensing TX (part of gNB 1) transmitting the sensing signal, which is scattered/reflected and received by multiple RXs. This may also involve a combination of a (quasi-)monostatic and a bistatic transmission where the reflected TX signal may be received by one RX on the same site and by additional RXs at neighboring sites. The signaling follows the same principle as in Figure 2, though here it forms a joint procedure in which no sensing occurs if configuration of either TX or RX gNB fails. Like in multi-monostatic, multiple concurrent multi-bistatic sensing operations are possible, i.e., there can be multiple TXs transmitting with overlapping coverage, if orthogonal TX signals are configured (otherwise only receive angles but no range can be determined).

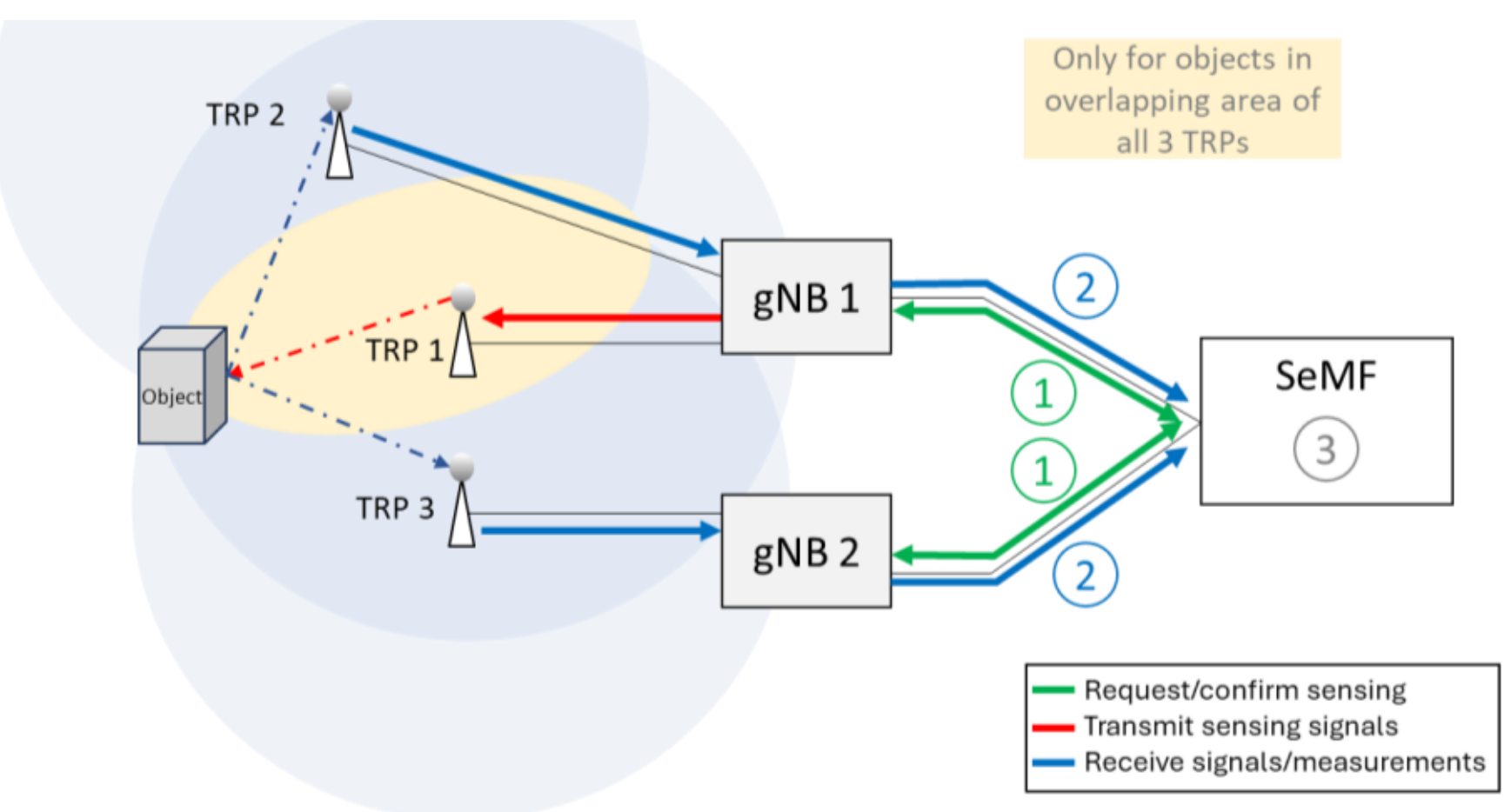

*Figure 3: Multistatic (multi-bistatic) sensing.*

## RAN Logical Architecture

Enhancing the RAN for sensing requires processing of the reflected TX (transmit) signals received on the same frequency. The logical RAN architecture with ISaC enhancements is shown in Figure 4. Existing communication functionality is maintained as is. The RU connects via the OpenFH (O-RAN Open Fronthaul [7]) interface to the digital processing part of the RAN. For sensing the OpenFH transport of frequency domain IQ (In-phase Quadrature) samples is reused, at least for OFDM (Orthogonal Frequency Division Multiplexing) radar like sensing (see section Sensing Signal Processing below). Should future OpenFH specifications evolve towards more processing in the RU, moving part of "L1-high TX/RX" into "RU" and for example switch from transporting frequency domain IQ samples per antenna port to, e.g., (quantized/noisy) QAM (Quadrature Amplitude Modulation) symbols per spatial layer, sensing then likewise should move part of "L1-high Sensing" into the RU and switch from IQ samples to, e.g., transporting amplitude-phase samples in delay-Doppler-angular domain. Specifically for large mMIMO (massive Multiple Input Multiple Output) arrays the fronthaul link bandwidth limits the number of possible spatial IQ sample streams and such a switch could help in exploiting their full spatial capability without incurring a large increase in fronthaul data rate.

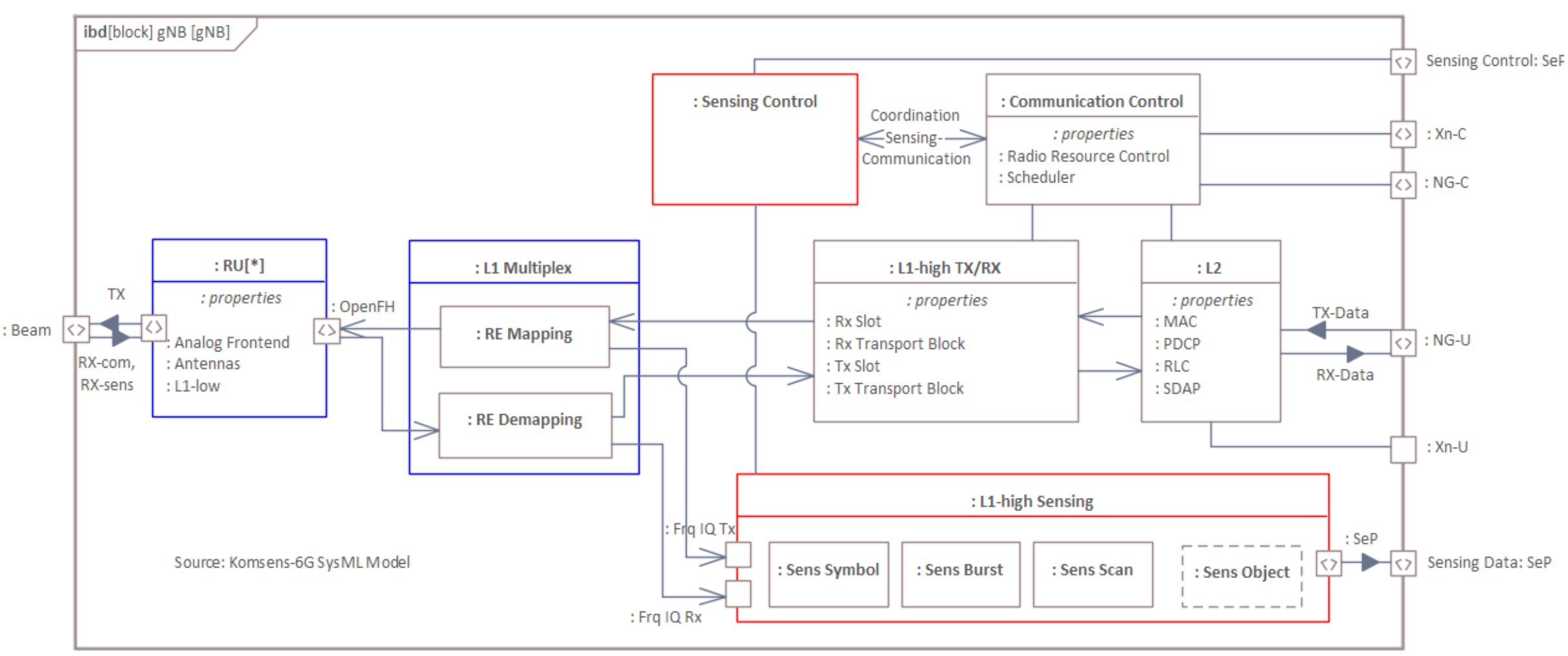

*Figure 4: RAN Architecture for ISaC.*

## Control Plane

The gNB architecture shown in Figure 4 includes the new sensing control function "Sensing Control". The sensing functionality of the RAN is requested and controlled by the SeMF via SeP over the sensing control interface (one of the proposed options in Section Logical System Architecture for ISaC). Figure 5 illustrates how the SeMF can signal a sensing request to a pair of TX- and RX-gNBs. Alternatively, the SeMF may signal a sensing request to a TX-gNB that then selects and signals a sensing request to a suitable RX-gNB or to a RAN node that selects and signals a sensing request to both a TX-and RX-gNB.

Sensing processing, irrespectively of requested sensing result, follows the same processing as communication up to "L1-high Sensing", wherein the sensing data are processed, based on the requested sensing result, and from which all requested sensing data are then extracted and forwarded via SeP to the SeMF.

The "Radio Resource Control" (RRC) is an established protocol running between the gNB and the UE. In the context of sensing it is involved when UEs are acting as sensing transmitter or receiver, which is out of scope for KOMSENS-6G.

The "Scheduler" performs dynamic allocation of radio resources among all active users and services, multiplexing the limited resources among them. In doing so, the scheduler needs ways to trade off the performance between sensing and communication demand. For sensing, the resource allocation requirements depend on the SQoS parameters provided by the SeMF to the Sensing Control block. If and to what extent the SeMF may have a role in influencing the resource allocation is for further study. The general principle of functional split between core and RAN is that the latter is responsible for the radio resource allocation.

For ISaC we can distinguish these options:

1) Communication and sensing each use their own signals and are multiplexed.
2) Communication signals are reused as sensing signals.
3) Communication signals are reused for sensing opportunistically where this is possible, and sensing signals are inserted where reusing communication signals is insufficient. This can be regarded as a combination of case 1) and case 2).

The scheduler can make use of these options, also depending on policies or priorities between communication and sensing.

A resource allocation pattern that is uniform—in particular periodic—in time and frequency, has advantages regarding processing complexity. The Sensing Control can aggregate the requirements from multiple sensing requests and may be able to fulfill it by a more efficient (smaller) resource allocation than the sum of the allocations for serving each request individually.

In TDD (Time Division Duplex) systems sensing transmissions would be preferably scheduled in downlink slots. In bi-/multistatic sensing, the receiving TRPs may belong to different gNBs. In this case the sensing schedule needs to be signaled also to the RX-gNB as shown in Figure 5.

Sensing resource allocation is preferably periodic. Such preconfigured sensing signals as in case 1) in the list above have the advantage that the configuration information completely defines the sensing signal for the receiver. Note here that gNBs make the final decision on scheduling resources for both communication and sensing. This is reflected by the Sensing Control block which should be able to communicate with other gNBs if cooperation in terms of resource allocation is needed, e.g., bi-static, or multi-static sensing.

Instead of relying on dedicated preconfigured sensing signals, in case 2) downlink communications signals are reused. As the RX-gNB does not know the complete TX signal as reference in advance, it must receive it from the TX-gNB either directly over the air, typically over a LOS (Line-of-Sight) path, or via backhaul as depicted in Figure 5 (in monostatic sensing the TX signal is readily available as it is the same gNB).

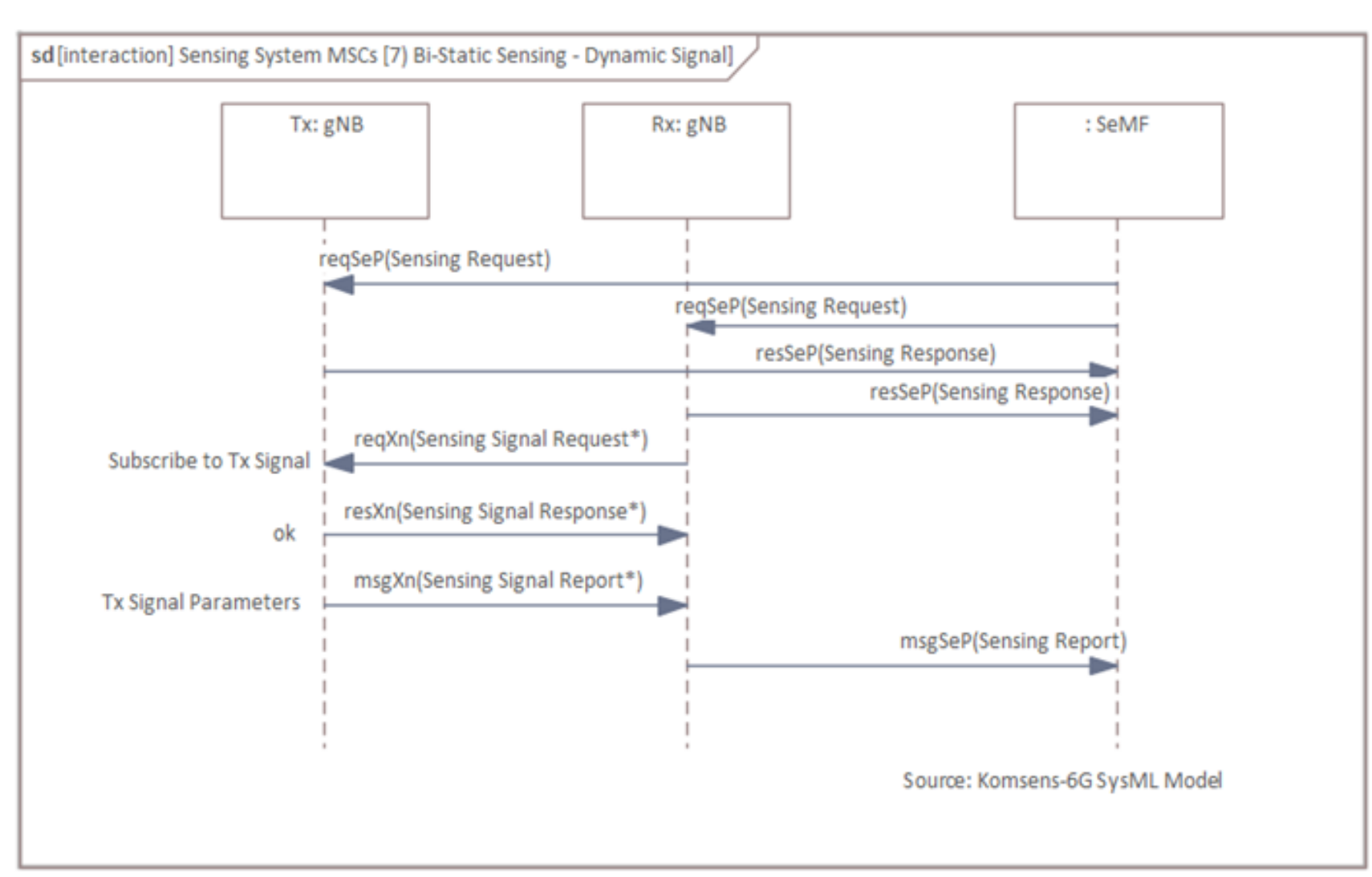

*Figure 5: Bistatic sensing on dynamic signals.*

Reception of the reference signal over the air only works with communication signals using wideband precoding (as any per sub-band precoding weights cannot be inferred from the received signal), and decoding may fail, depending on SINR (Signal-to-Noise and Interference Ratio) in relation to the MCS (Modulation and Coding Scheme) used for the communication signal. Transfer of the reference signal via backhaul, e.g., as shown in Figure 5, the RX-gNB requesting the TX-gNB to forward TX signals matching the RX-gNB's own sensing schedule, requires a sufficiently high bandwidth interconnection for conveying the TX signal, e.g., as frequency domain IQ samples, at the time of transmission, and with the interconnection being sufficiently low latency or the sensing receiver having sufficient receive IQ sample buffer space for buffering the received (sensing) signal until the TX signal has arrived over backhaul.

A simpler approach to case 2) is therefore to exploit only the communication reference signals that are embedded in the complete communications signal for sensing, e.g., DMRS or CSI-RS. In this variant the RX-gNB can be preconfigured with these reference signals and in case of backhaul signaling only an index to them needs to be dynamically signaled. Autocorrelation properties of these RS are also better than those of the data signal that is embedded in the communications signal.

Since communication is predominantly scheduled dynamically, reusing communication signals for sensing is opportunistic in nature. At times when there is no ongoing communication in direction of interest for sensing, a sensing signal needs to be transmitted to fulfill the sensing request, leading to case 3). The overall sensing schedule, e.g., in terms of how frequent a measurement result needs to be delivered may still be determined by the SeMF and the sensing TX-gNB scheduler can then decide whether it sends a communication signal or a known sensing reference signal on a scheduled sensing resource. In bi-/multistatic sensing, the RX-gNBs could receive the sensing schedule from the SeMF and their schedulers would have to switch to RX mode in the sensing resources and would have to schedule their communication around them.

If the sensing transmission and reception is supposed not to follow a semi-statically configured schedule, dynamic scheduling flexibility can be provided. While this is no issue in monostatic sensing, in bi-/multistatic sensing the scheduler in the TX-gNB must inform the RX-gNBs with sufficient time ahead due to inherent delay between two gNBs to switch to sensing RX mode. Since this typically has an impact on the RX scheduler, the dynamic scheduling could imply some brief handshake between the TX-gNB and RX-gNB before the actual sensing transmission starts. Such delay-critical signaling between TX-gNB and RX-gNB needs to use a direct RAN-internal low-delay interface like a 6G evolution of the Xn interface or a new interface or a link to a control unit in the RAN.

Since sensing is supposed to happen in DL-symbols, the sensing receiver may be interfered by transmissions from other TRPs for communications or sensing purpose. Such TRP to TRP interference is often called cross link interference (CLI). In order to mitigate CLI, coordination between TRPs is required. The coordination may imply restrictions on the interfering TRP such as reducing transmit power, avoiding certain symbols, subbands or beams. In addition, certain beamforming techniques may have to be employed at the interfering TRPs, like creating a notch in the beam pattern towards the interfered TRPs. If the interfering signal is itself not bearing communication information but used for sensing, then it can also be tailored in a way that the interference appears just as if there was a sensing target at the location of the interfering TRP, which could be filtered out at the interfered TRP. This can be achieved e.g., by using signal sequences for the desired and the interfering sensing signal that are identical except for a possible cyclic shift in the delay or Doppler domain.

With TRPs mounted above roof-tops, strong CLI can appear over long distances, typically 10s of km, and under rare adverse weather conditions (ducting) even 100s of km. The coordination between TRPs therefore needs to extend over more than just the adjacent TRPs. The radio resource coordination should be self-contained in the RAN.

**Sensing Signal Processing**

The gNB includes a new component "L1-high Sensing" for the digital processing of the sensing signal. Figure 6 outlines the processing steps of "L1-high Sensing" in case of an OFDM based radar [8]: "Sens Symbol" operates on an OFDM symbol level, providing range information, "Sens Burst" on a burst of OFDM symbols additionally providing speed information, "Sens Scan" on a set of bursts, typically from an angular scan over a set of beams, providing angular information and spatial filtering, and the final "Sens Object" provides semantic interpretation of detected targets including tracking, aka filtering over time. In the following, the contained individual processing steps are explained in more detail.

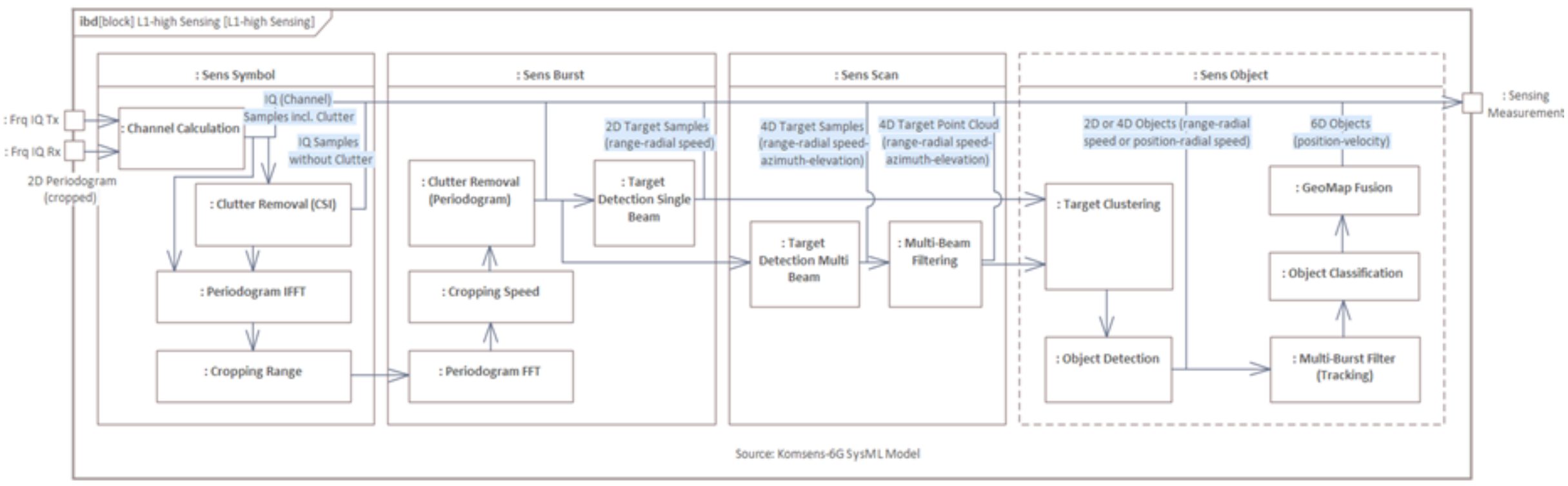

*Figure 6: Architecture of "L1-high Sensing" with processing options.*

“Channel Calculation” calculates the effective radio channel, independent of the actual transmitted sensing signal, to be captured in the output IQ (Inphase Quadrature) samples. To reduce computational effort, the input IQ samples may be decimated in frequency and/or time, i.e., only every kth subcarrier and only every lth OFDM symbol is used as input, reducing the unambiguous value range as well as radar processing gain (aka sensitivity). In contrast, bandwidth and/or burst duration determines the achievable resolution, which is often just marginally sufficient. Thus, as wideband and as long duration sensing bursts as necessary for the considered use case and that can be still coherently transmitted and received, are preferred.

From the effective radio channel, “Clutter Removal (CSI)” removes major channel components identified as being clutter, i.e., as being channel components of no interest. The clutter removal also enables detection of weak targets that otherwise may be masked by possibly stronger clutter channel components. It is an optional feature that can be skipped, e.g., when the (semi-)static environment is to be sensed. Alternatively, or additionally, clutter removal can be carried out after periodogram computation, denoted by “Clutter Removal (Periodogram)”.

Next, “Periodogram iFFT” and “Periodogram FFT” compute a complex-valued 2D periodogram, i.e., a complex-valued 2D matrix with range and radial or bistatic speed (calculated from delay and Doppler) as dimensions. “Cropping Range” and “Cropping Speed” right after the (i)FFT of the respective dimension may be used to reduce further processing to the area and speed range of interest, reducing both memory and processing requirement of the following steps.

The “Target Detection Single Beam” extracts targets by searching for local power maxima within the periodogram. Up to this point, a single sensing burst, i.e., a (continuous) coherent receive signal, is used as input. Alternatively, multiple sensing bursts from different receive beams are collected, for each a range-speed periodogram is computed, which are then stacked in order of their respective beam’s azimuth and/or elevation angle to create an up to 4-dimensional periodogram tensor. “Target Detection Multi-Beam” searches for local power maxima therein and “Multi-Beam Filtering” filters the detected targets in space, using multiple sensing bursts from different receive beams to provide a location estimate for each target in form of an estimated azimuth (AoA) and zenith (ZoA) angle of arrival in addition to the estimated range and speed.

Before filtering in time, “Target Clustering” and “Object Detection”, which work in tandem, need to identify which target points in the 2D or 4D point cloud stem from the same object as respective target points in prior point clouds, possibly taking into account prior results from “Multi-Burst Filtering” to improve correct mapping of target points to objects. They group multiple targets, each group representing a single sensed object, if the targets exhibit similar velocity and maintain proximity over time and, thus, seem to stem from a single physical object like a human or car. For each such detected object, “Multi-Burst Filtering (Tracking)” updates each of the detected object’s estimated position, velocity and possibly orientation (from relative movement of the object’s constituent target points) in 2D or 3D space with the current sensing sample. Specifically, the direction of travel is determined. “Object Classification” tries to determine the probability of object (human, car, truck, drone, bird, tree, building, etc.). “Geomap Fusion” with help of geolocation information provides further context, e.g., for removing ghost targets, i.e., targets created by reflections from known buildings, or improbable direction of travel, and helping in classification of objects.

The spatial filtering “Sens Scan” may be skipped, e.g., in case of simple TRPs like small cells that typically lack beam steering capability, or when the use case does not require it, or when timely reporting of sensing measurements is more important. 2D targets may then be reported as is, or “Sens Object Detection” may identify and possibly group targets based on proximity, specifically if subsequent filtering in time is to be done.

Not all the above processing steps need to be performed all the time and/or on all parts of the periodogram and/or for all targets, depending on how “ISaC-Control” has configured “L1-high Sensing”. Channel estimation, all symbol-level processing (“Sens Symbol”) or even whole periodograms (“Sens Burst”) may have been done already in the RU. The object detection and tracking can be done either in gNB or left to the SeMF as indicated with “Sens Object” being shown as dashed block in Figure 6 and being included in SePF part of the SeMF in Figure 10. The gNB then reports, possibly spatially filtered by “Sens Scan”, target point clouds to the SeMF. Specifically, if “L1-high Sensing” processing stops prior to “Target Detection”, the created high volume periodogram data or IQ sample stream motivates a dedicated DP (Data Plane) or interface for forwarding such data and possibly a part of the sensing data processing and fusion happening in a processing node similar to the SePF co-located with the RAN.

In an edge could deployment, some or all of the processing steps may be realized as one or more VNFs (Virtual Network Function), specifically target and object detection, and spatial and temporal filtering. Furthermore, all presented functions may be partly or fully realized through AI/ML (Artificial Intelligence/Machine Learning), e.g., as a DNN (Deep Neural Network) that inputs IQ samples and outputs 2D or 4D target point clouds, or even readily identified and tracked objects.

**TRP and RU**

For the RU the main challenge is the interference of concurrent strong RX (mostly communication) signals with the low power RX (received sensing) signals. Three different RU architectures are considered to cover different deployments: TDD as shown in shown in Figure 7, FDD (Frequency Division Duplex) as shown in Figure 8 and “Sniffer” (separate RX-only TRP) as shown in Figure 9, each adding the necessary functions (marked in red) for sensing. When receiver and transmitter are co-located, there is high self-interference, thus, for SIC (Self-Interference Cancellation) following options are proposed:

- antenna isolation and/or shielding of RX from TX antenna array,
- analog signal cancellation on RF (radio) or BB (baseband) frequency, i.e., prior or after RF down-conversion,
- digital signal cancellation in time or frequency domain, i.e., prior or after FFT (Fast Fourier Transform), allowing better signal compression for transport over the OpenFH interface.

In bi-/multistatic sensing, self-interference can be avoided by pausing all communications in the same band on RUs acting as sensing receivers with the drawback of increasing the overhead of sensing, but with the advantage that legacy TDD RUs can be used as sensing receivers without modifications.

The impact of interference from adjacent carriers and neighboring cells still needs to be analyzed.

*RU for TDD*

Figure 7 depicts the internal architecture of an ISaC-capable TDD RU. All newly introduced functions for sensing are highlighted in red. In TDD operation, when a TRP acts as sensing transmitter, the same RX path can be used for both sensing and communication: During the uplink (UL) part of the TDD period, the RX path is used for communication, and during the downlink (DL) part it is used for sensing. The “s: RX Array” (s: sensing) used solely for sensing may be realized as part of the “trx: TX/RX Array” (trx: transmit and receive) for communication, which is reconfigured for sensing such that one part acts as TX array for transmitting sensing signals and user data, and another part as RX array for receiving sensing signals. The RU may solely employ analog beamforming (BF) as it is typical for mm-wave frequencies as in frequency band FR2. An RU with hybrid or fully digital beamforming typical for cm-wave frequencies as in frequency band FR1 and possible future bands additionally or exclusively has a second respectively its only beamforming function within “L1-low TDD” before the iFFT and after the FFT, respectively. “L1-low TDD” must have enough processing power for processing received signals, not only during the UL part but additionally during the DL part of the TDD period.

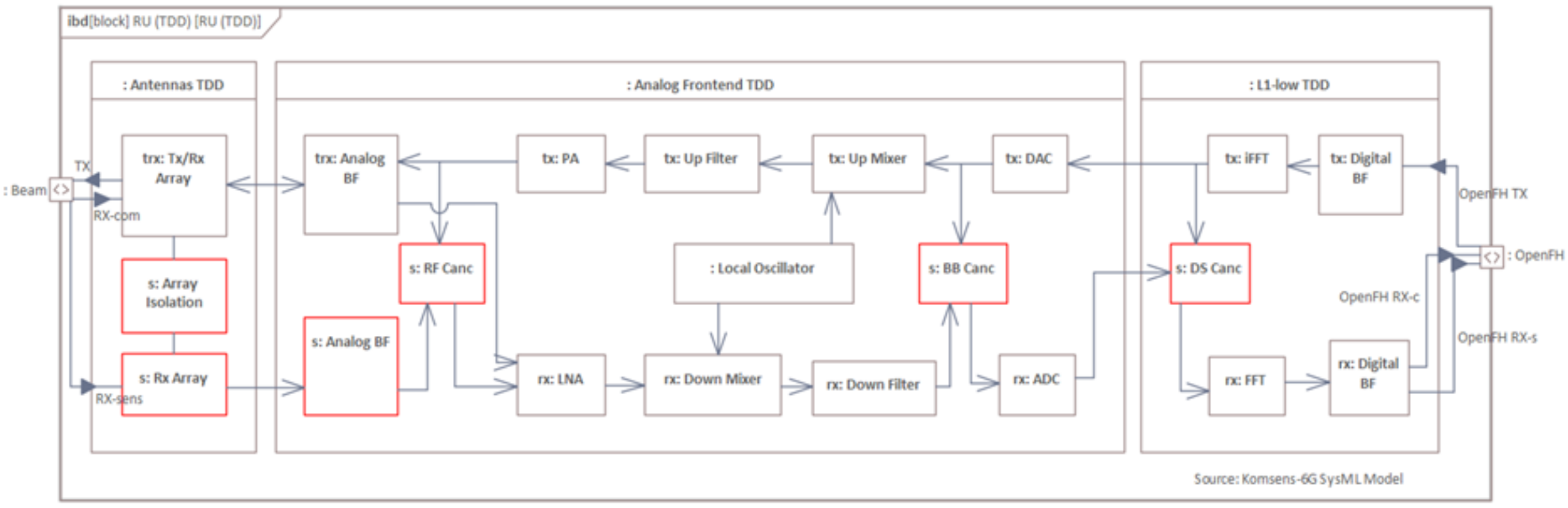

*Figure 7: ISaC Radio Unit Architecture for TDD.*

While the above introduced means of self-interference mitigation (in red) are strictly necessary only for monostatic sensing, bi-/multistatic sensing also benefits as communication on the RUs acting as sensing receiver then can continue during sensing.

*RU for FDD*

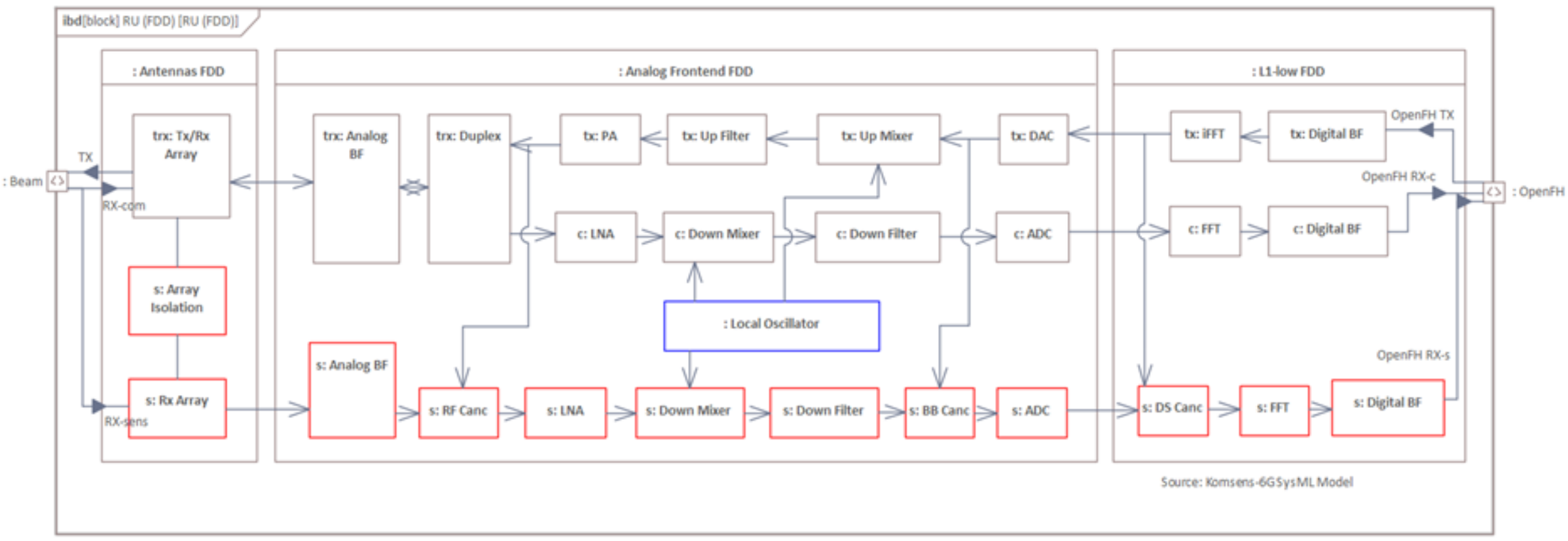

*Figure 8: ISaC Radio Unit Architecture for FDD.*

For FDD, as shown in Figure 8, the logical architecture of the RU comprises an additional RX path to enable monostatic (both TX and RX in one RU), as well as bistatic (TX and RX in different RUs) sensing. While both RX paths, communication (c) and sensing (s), comprise logical components of the same type, the communication path operates on UL frequencies, while the sensing RX path operates on DL frequencies, the same frequency as the TX path. Accordingly, the same self-interference mitigation techniques as for TDD are needed for the sensing RX path (but not for the communication RX path)—except the RU is only used for bi-/multistatic sensing and during sensing reception all communication in the same band is paused.

*Sniffer RU*

When an RU is used only for receiving (sensing) signals, it can be operated in ‘sniffer’ mode, see Figure 9. The sniffer RU may be a standard TDD RU with its TX path deactivated, or the RU is purpose-build without TX path (specifically in case of FDD). As the TX path is either inactive or even missing, interference cancellation is not required. The "Sniffer" is applicable to a quasi-monostatic setup, i.e., where a second co-located TRP acts as sensing transmitter, and where the sniffer RU respectively its RX antenna can be placed such that the required isolation from the TX antenna of the co-located TRP is achieved. To achieve similar performance as an integrated sensing RU, the local oscillators of both the transmitting (non-sensing) RU and receiving Sniffer RU may use a common reference oscillator.

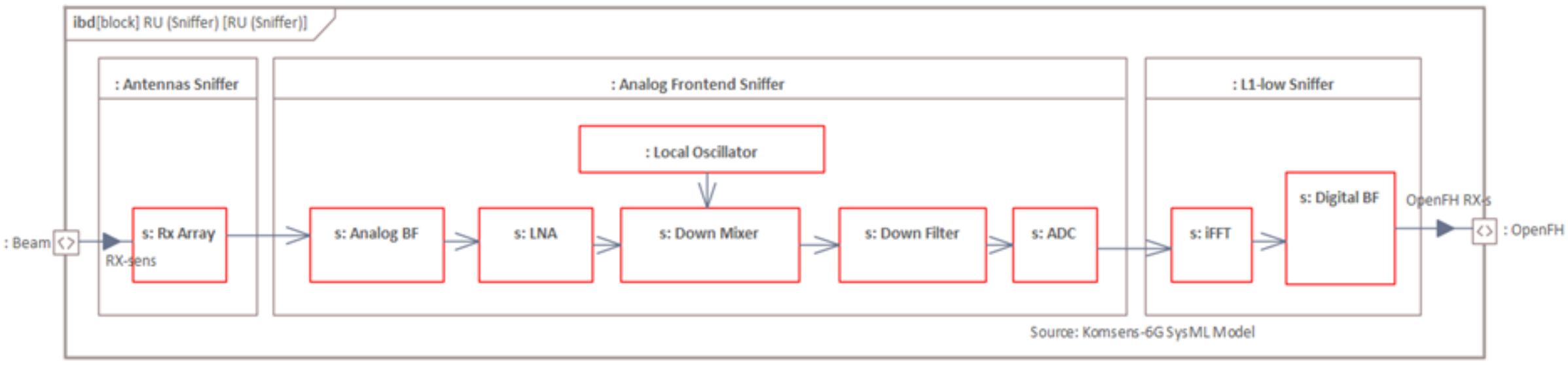

*Figure 9: 'Sniffer' (receive only) ISaC Radio Unit Architecture.*

#### Core Function SeMF

The "Sensing Management Function" is added to the core functions to anchor all new functions required for the integration of sensing, see Figure 10. It offers its services to other core functions via the Nsemf interface.

The SeMF provides the "ISaC-API" to expose the sensing service to external applications. The SeMF controls the RAN sensing functions via the SeP. Both, ISaC-API and SeP communication interfaces are part of the architecture proposal.

The SeMF (Sensing Management Function) contains several subfunctions, namely:

- **SeCF (Sensing Control Function)** manages all network actions for any Sensing Trigger received from the API. It may include:
    - **TRP Discovery and Selection** discovers available TRPs.
    - **SQT (Sensing Quality Translation)** to translate the input SQoS requirements into requirements on resources, power, sensing mode, etc.
    - **Sensing Mode Management** decides on the mode a) monostatic, b) multi-monostatic, c) multistatic (incl. bistatic) and requests sensing measurements from selected TRPs for sensing in the requested area as needed.
    - **PeP (Policy Enforcement Point)** enforces privacy-related policies within SeCF.
- **SePC (Sensing Policy Control) Function** is a logical collection of privacy-enabling functions including
    - **SPCTM (Sensing Policy, Consent and Transparency Management)** the central policy decision point, governing access to sensing data based on defined policies, user consent, and regulatory requirements like GDPR.
    - **CAF (Consent Acquisition Function)** attains consent from sensing targets.
    - **TEF (Transparency Exposure Function)** ensures necessary transparency information is provided to individuals involved in the sensing.
    - **PIF (Privacy Imposing Function)** determines the specific technical privacy controls to be applied, such as data minimization or anonymization.
- **SePF (Sensing Processing Function)** processes the received sensing data according to the request. It includes:
    - **Sense Object**, creating a list of objects from the sensing data. It includes "Target-Clustering", "Object-Detection", "Multi-Burst-Filtering" and the "Object Classification" identifying the type of object (building, car, human, etc.). As explained before these functions can be alternatively located in the RAN.

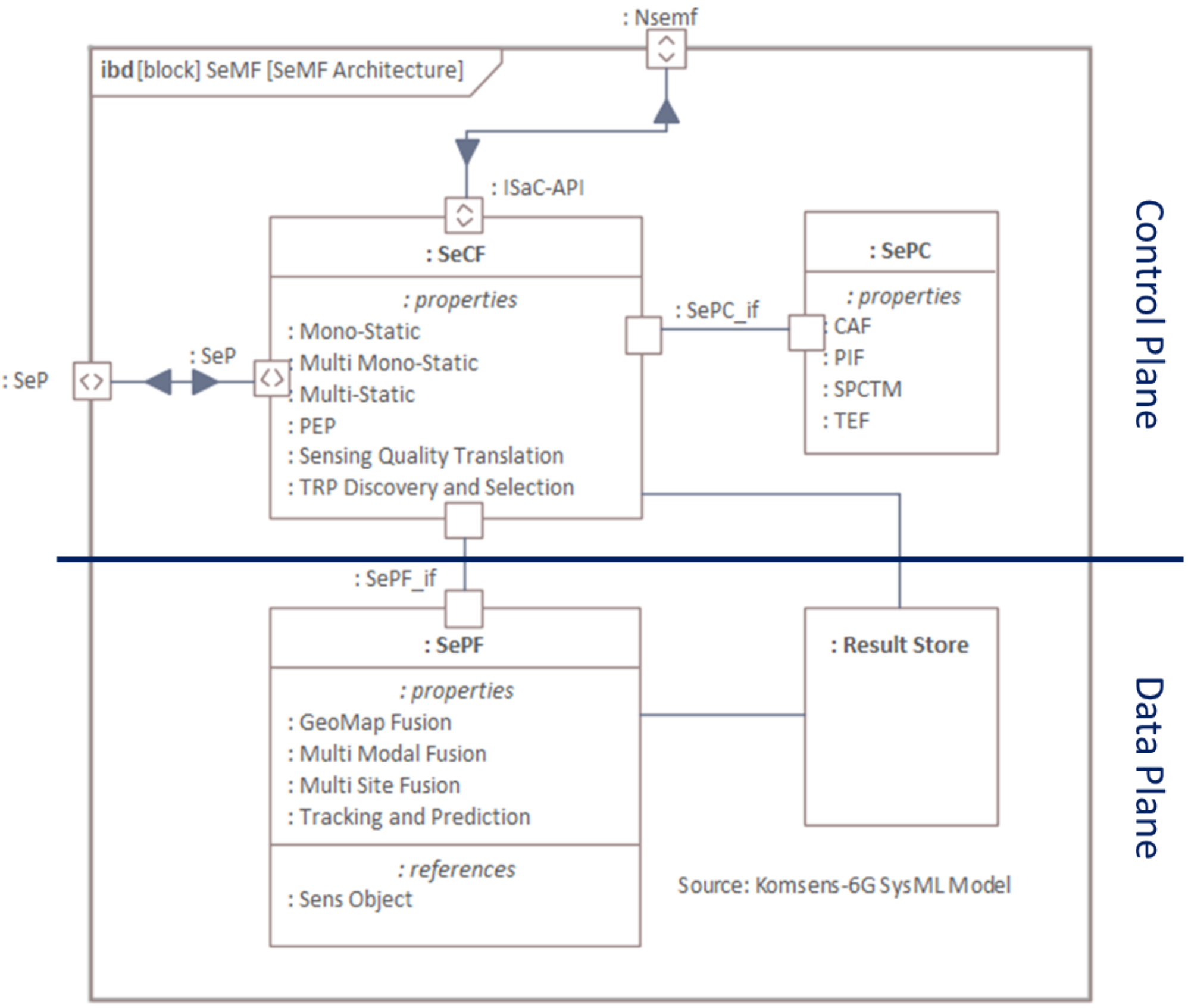

*Figure 10: Core Architecture of the Sensing Management Function (SeMF).*

- **Geo-Map Fusion** uses existing 3D Geo-Map information (e.g., CityGML[2]) to identify and classify sensing objects.
- **Multi Modal Fusion** fuses the sensing information with other sources (e.g., Lidar, video).
- **Multi Site Fusion** uses sensing information from multiple TRPs to obtain the requested sensing result.
- **Tracking and Prediction** identifies moving objects, tracks and predicts object movements.

- **Result Store** stores object locations and updates with every scan. Includes results from data processing before passing it on to the application function (AF). The Result Store can also serve multiple API-requests for the same area w/o initiating a sensing scan for each request.

Selected subfunctions like SePF and related control functions, may become separate entities and be co-located with the SeMF or be a part of the RAN. In particular subfunctions dealing with radio parameter configuration or direct sensing processing, like Sense Object, Multi Site Fusion, Tracking and Prediction, TRP Discovery and Selection, Sensing Quality Translation and Sensing Mode Management can make sense to be part of the RAN. Keeping such radio-near subfunctions in the RAN can also foster future innovation with no RAN-Core interface involved and keep operations and management systems and competences focused to the domains of RAN and Core, respectively.

### QoS for Sensing Aspects

QoS assurance plays a crucial role in enabling predictable and reliable operation of ISaC systems. Within the KOMSENS-6G framework, QoS management becomes particularly challenging due to the need to jointly handle sensing and communication demands that share the same physical and logical network resources. Sensing applications may require high accuracy or frequent updates, while communication services emphasize low latency or high throughput. Ensuring that both dimensions are satisfied simultaneously requires a coordinated and transparent approach to QoS (respectively SQoS for sensing) handling. In the current KOMSENS-6G architectural concept, applications express their service requirements toward the network through the ISaC-API, which interfaces with the SeMF. These requirements are typically formulated in high-level QoS categories (e.g., 3GPP-defined service categories). However, a gap remains between this abstract service intent and the detailed resource-level configurations that must be applied in the RAN through the SeP. To bridge this gap, the concept of a Sensing Quality Translation (SQT) has been introduced as a potential architectural enhancement. The SQT acts as an intermediary mechanism responsible for interpreting application-level SQoS intentions and aligning them with the resource management capabilities of the network. To support this translation, the SQT relies on a SQoS mapping table that defines the correspondence between application-level SQoS intents and the estimated amount of RAN resources required to meet those intents. This mapping ensures that service-level expectations are consistently converted into quantifiable network actions. At a conceptual level, it translates high-level service objectives into operational policies that the SeMF and RAN can enforce. This translation process is key to achieving consistent QoS behavior across the sensing and communication domains, ensuring that service expectations are effectively reflected in network operation.

---

[2] The CityGML standard defines a conceptual model and exchange format for the representation, storage and exchange of virtual 3D city models – https://www.ogc.org/standards/citygml/.

**Security and Privacy Aspects**

In 6G ISaC architectures, security and privacy controls are essential, particularly when sensing targets involve human beings. Such scenarios trigger data protection regulations like the GDPR, as the sensing data constitutes personal data. The SPCTM (Sensing Policy, Consent & Transparency Management) function is the central component for governing access to this sensitive data. It manages the sensing policies, which dictate aspects such as who is allowed to access which data, for what purpose, and with what frequency. Policy decisions are integrated with consent management, determining if sensing is permissible at a given location based on user or property owner consent.

For example, SPCTM can decide whether sensing is allowed based on attributes such as the time of the day, area requested for sensing, and anticipated resolution (e.g., in the 2D image space) of the post-processed sensing result. To illustrate this attribute-based governance, consider a scenario where a municipal traffic monitoring application requests high-resolution radar data at a busy intersection during peak commuting hours. The SPCTM might automatically enforce a policy that restricts the sensing resolution to a coarse point cloud—sufficient for counting vehicles but insufficient for facial recognition or fine-grained tracking—thereby dynamically balancing service utility with privacy preservation.

In addition to policies dependent on technical attributes, the SPCTM specifies the consent and transparency requirements for a given sensing request. Based on this, the SeCF (Sensing Control Function) triggers other specialized functions. The CAF (Consent Acquisition Function) is responsible for actively attaining consent from relevant entities. This function must handle complex scenarios, such as obtaining consent from users connected to different MNOs (Mobile Network Operator) or even users not connected to any MNO, ensuring regulatory compliance across diverse user states. Simultaneously, the TEF (Transparency Exposure Function) manages the dissemination of all necessary information to users involved in the sensing. The TEF may utilize mechanisms like broadcast messages or public disclosure logs to inform targets about the sensing purpose, retention period, and risks, ensuring the “right to be informed” is respected.

Furthermore, the SePC facilitates data minimization, ensuring that applications only receive the minimal data necessary for their purpose. To strictly enforce this, the PIF (Privacy Imposing Function) decides on specific technical controls to be imposed during both the sensing and data processing stages. A critical aspect of minimization is the application of noise addition techniques, such as Differential Privacy or Local Differential Privacy, depending on the trust model. For instance, if the sensing data is processed centrally in the SePF, standard Differential Privacy mechanisms add calibrated noise to the processed sensing result. Conversely, if processing occurs closer to the edge or user (e.g., RU or gNB), Local Differential Privacy ensures noise is added before the data leaves such a trusted boundary.

These controls are vital for satisfying the privacy requirements of unlinkability and unobservability, concepts central to privacy engineering standards like ISO/IEC 15408-2 (Common Criteria). Unlinkability ensures that a user cannot be distinguished from others within the set of sensing subjects, nor can their actions be correlated across different sensing sessions. Unobservability ensures that a user may use a resource or service without others, especially third parties, being able to observe that the resource or service is being used. By mandating noise injection via the PIF, the architecture ensures that even if sensing data is intercepted, the PII (Personally Identifiable Information) within it cannot be linked back to a specific individual, thereby preserving anonymity.

Finally, the SePC interacts with the PEP (Policy Enforcement Point) located within the SeCF via the SePC_if interface. Through this interaction, SeCF enforces the decisions made by SePC's functions. This enforcement extends to data processing; for instance, the SeCF instructs the SePF on which privacy controls, decided by the PIF, must be applied to the sensing data. These operations are ideally conducted within ephemeral processing sessions to prevent long-term data retention risks.

### Signaling Flow and Protocols

A typical message flow for processing a sensing request from a sensing AF is shown in a sequence diagram in Figure 11. If the TRP information is not yet known, it is retrieved from the RAN. When the TRP information is available, a sensing request is passed to the gNB. The sequence can end by providing sensing results or by termination via API or gNB failure indication. Note that more message sequences are available for the SeP as well as for the ISaC-API, defining successful and unsuccessful operations in the KOMSENS-6G SysML Model but are not included in this whitepaper.

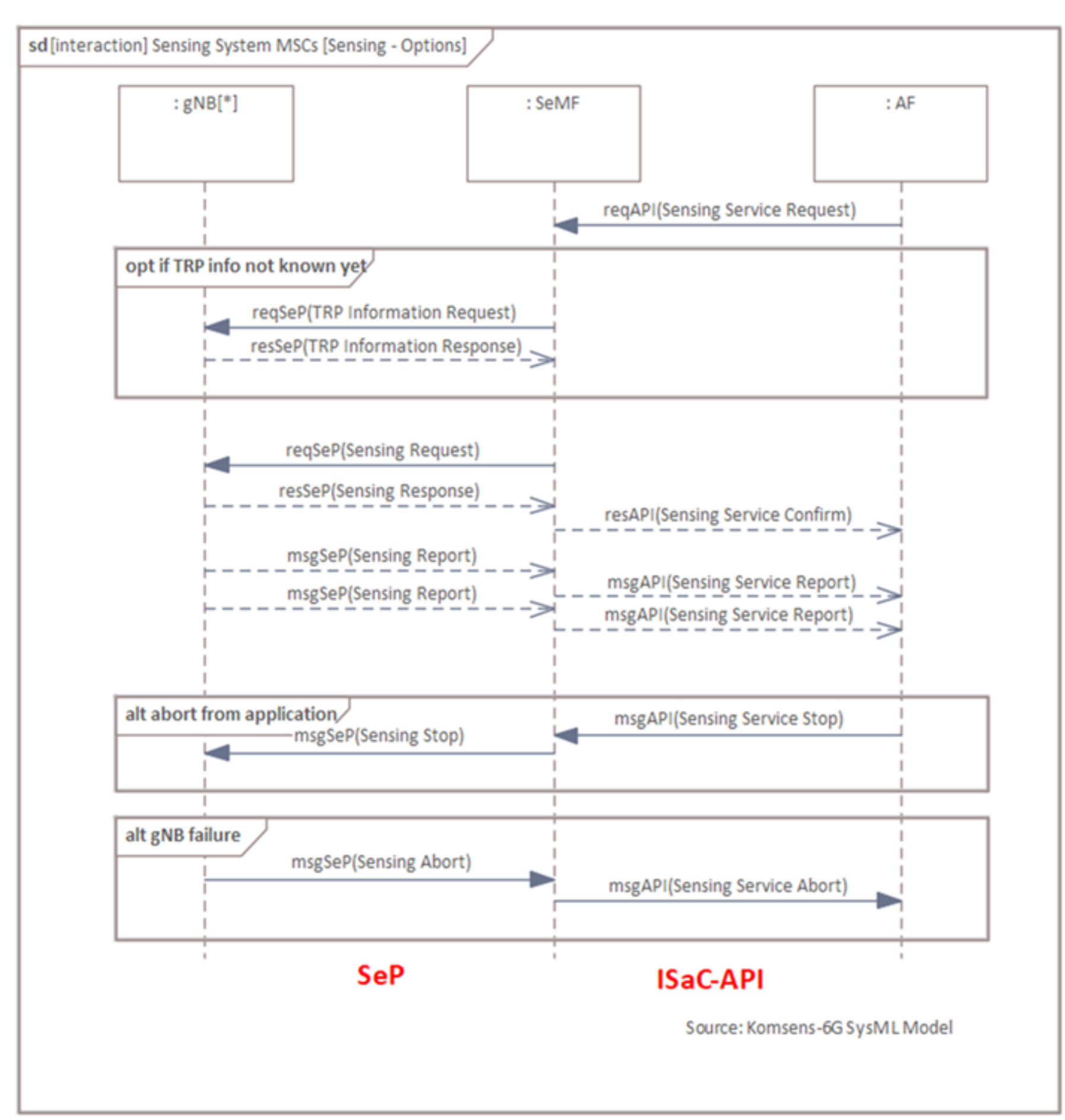

*Figure 11: Message Sequence of Sensing Service.*

*The SeP*

The "Sensing Protocol" (SeP) has been designed to enable the SeMF core function to control the sensing functions in the RAN. The SeMF uses SeP to request sensing-related information and measurements from gNBs. Table 3 summarizes all elementary procedures and Figure 12 shows the BDD (Block Definition Diagram) defining the set of their SeP messages and therein contained highest-level IEs (Information Element). Further details of the IEs are specified in the KOMSENS-6G SysML Model.

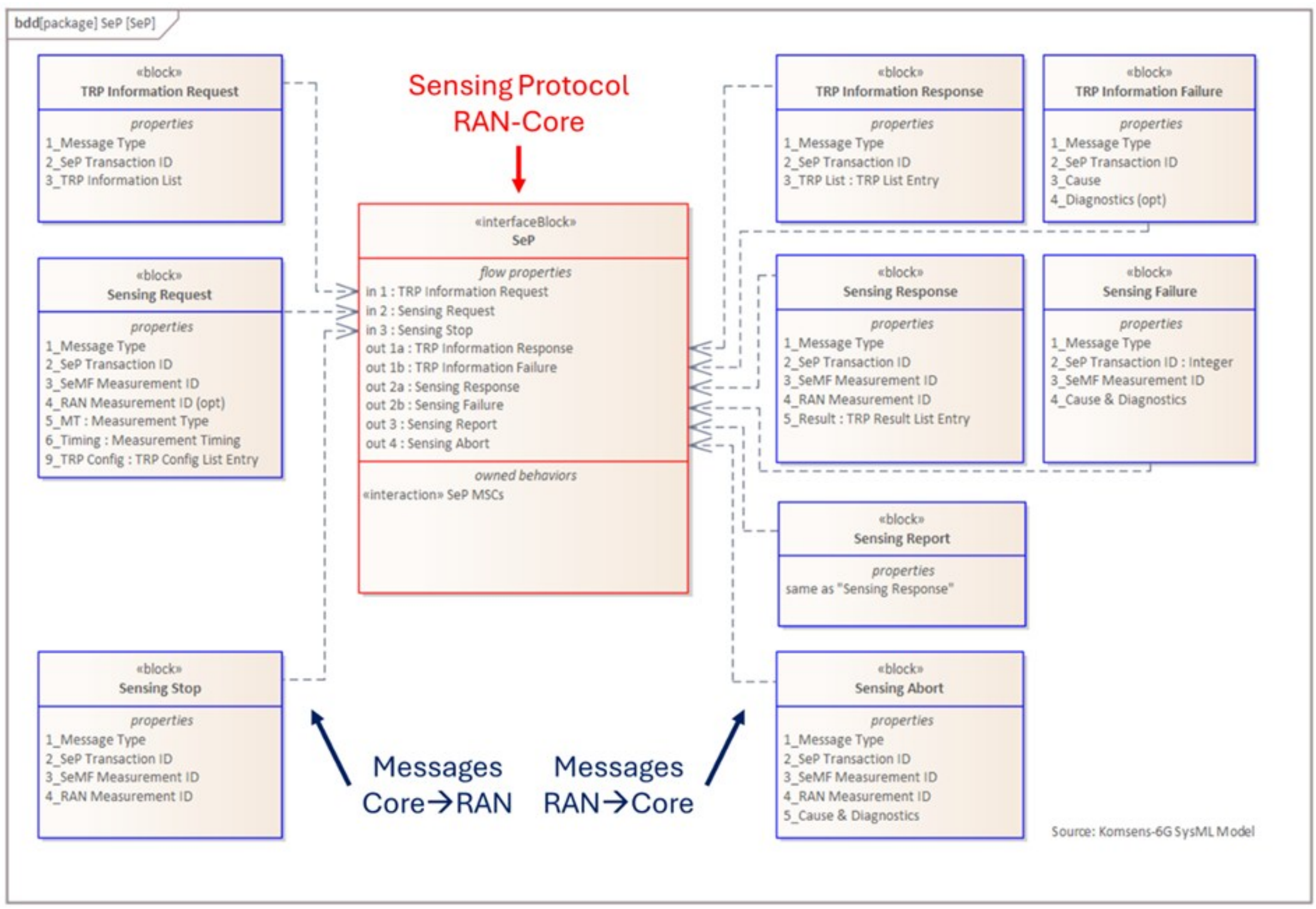

*Figure 12: Sensing Protocol (SeP) Messages and Parameters.*

SeP provides two functionalities, one for the SeMF to request information about TRP properties and capabilities (three messages, 1st row in Figure 12 and Table 3) and the other to carry out actual sensing measurement operations (all remaining messages).

*Table 3. Sensing Protocol (SeP) procedures for capability retrieval and sensing operation.*

| *Request-Response Procedures* | | | | *Single Message Procedures* |
|---|---|---|---|---|
| **Procedure Name** | **Initiating Message** | **Successful Outcome Response Message** | **Unsuccessful Outcome Response Message** | **Procedure Name/ Initiating Message** |
| TRP Information Exchange | TRP Information Request | TRP Information Response | TRP Information Failure | Sensing Report |
| Sensing | Sensing Request | Sensing Response | Sensing Failure | Sensing Stop |
| | | | | Sensing Abort |

With "Sensing Request" the SeMF activates sensing operation in the RAN with the configuration given in IEs "TRP Config List Entry". The contacted gNB either fully acknowledges the request with "Sensing Response", providing its own "RAN Measurement ID" for later reference and reporting of results, or (fully) rejects it with "Sensing Failure", then including "Cause & Diagnostics" providing the reason for the rejection. The case when a gNB may only be able to partially acknowledge a request with possibly modified measurement configuration(s) is still open for future study. When the SeMF has requested an immediate one-time (aka on demand) measurement in "Measurement Timing", and the result can be provided immediately (before the response timeout set out by the SeMF), "Sensing Response" already carries the sensing measurement results in IEs "TRP Result List Entry", and the sensing measurement operation is terminated with this message.

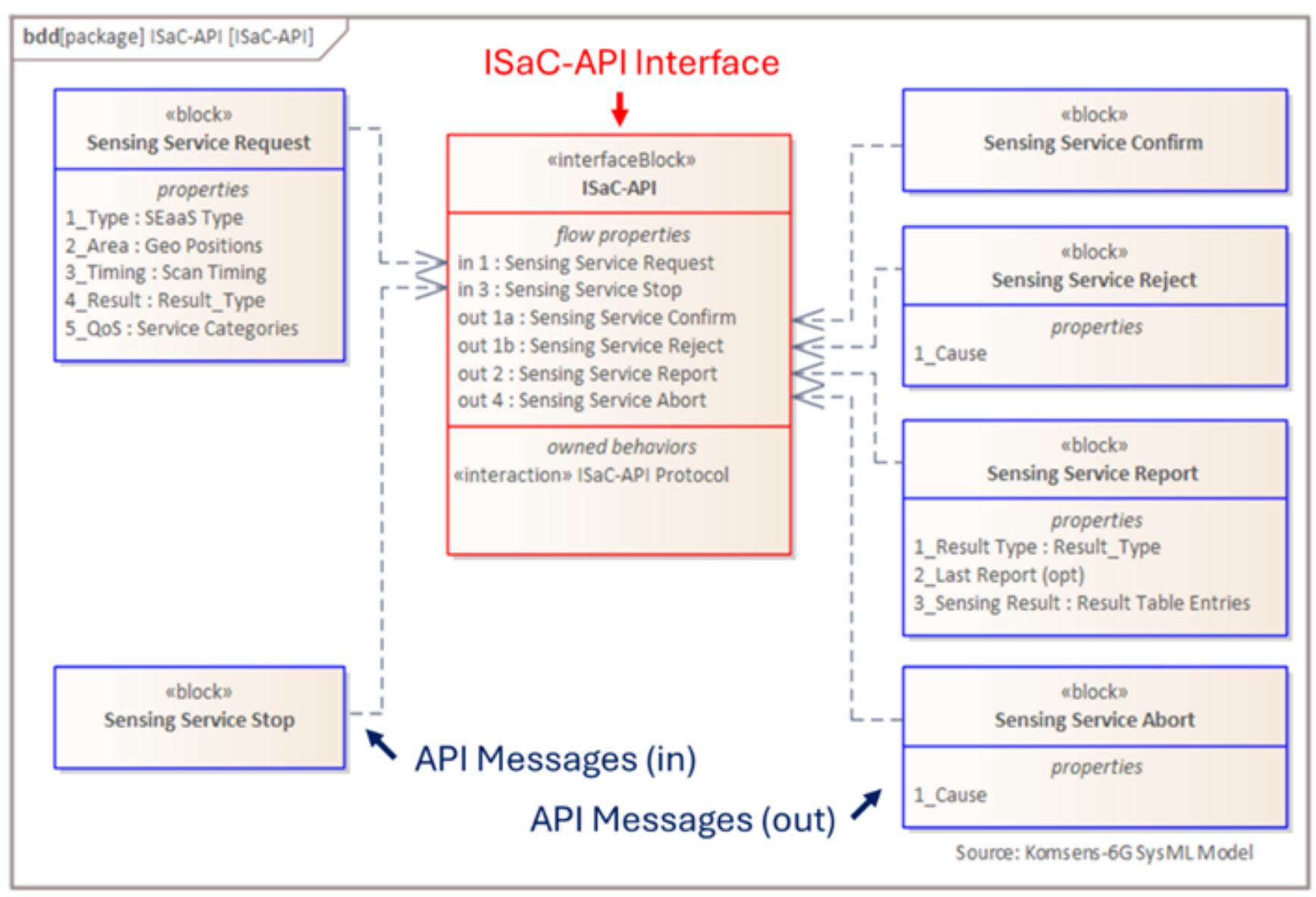

*Figure 13: ISaC-API Messages and Parameters.*

"Sensing Report" is used to provide for one-time measurement results that incur a longer delay like scanning the whole environment, as well as for periodic reporting of sensing measurements, either until the configured sensing duration (provided before in "Measurement Timing") has been reached or the sensing measurement operation is terminated by the SeMF with a "Sensing Stop". This message is also used in case a failure occurs on SeMF side. Conversely, "Sensing Abort" allows the RAN to abort an ongoing sensing measurement operation in case of a failure on RAN side.

By including the optional "RAN Measurement ID" within "Sensing Request", the SeMF can modify an ongoing (periodic) sensing measurement operation as identified by the tuple of "SeMF Measurement ID" and "RAN Measurement ID". In case the gNB rejects the change, the sensing continues as is and the SeMF may instead terminate it using "Sensing Stop" and request a new sensing measurement as described above.

*The ISaC-API*

The sensing service is exposed through an ISaC-API that has been defined within the KOMSENS-6G project. The ISaC-API (illustrated in Figure 13) defines functionalities such as requesting sensing services, results reporting and service termination as well as service abortion in case of failure. The general availability of the ISaC service in an area is expected to be exposed through other means than the ISaC-API itself. The ISaC-API has been checked against top use cases of KOMSENS-6G to validate the completeness of the functionalities.

**Summary**

The KOMSENS-6G project has identified a set of functions to realize an ISaC system in future 6G mobile communication systems. The proposed architecture is lean to minimize impact on other functionalities and implementation complexity. All core functions related to sensing are consolidated in a single new core network function named SeMF (Sensing Management Function). The sensing service is exposed via the ISaC-API provided by the SeMF. On the RAN side, baseband processing is amended by a second dedicated processing chain "L1-high Sensing" for sensing besides the existing "L1-high TX/RX" for communication services, and an extended "L1 Multiplex" resource (de)mapping for (de)multiplexing frequency IQ samples among both pipelines. In a classical distributed RAN with existing communication-only gNBs, "L1-high Sensing" may be initially realized as a separate processing function in the RAN, in a Cloud RAN deployment, by implementing most of the sensing-related processing as VNFs running on the edge cloud. "L1-high Sensing" is complemented by the new "Sensing Control" control layer function. It represents the RAN-side endpoint of the SeP (Sensing Protocol), by which the SeMF controls sensing operations in the RAN and receives the sensing measurements back from the RAN. For monostatic sensing a second RU in receive-only configuration ("Sniffer RU") can be used, which is physically separated from a co-located communication RU acting as sensing transmitter. For bi-static sensing communication transmissions in the same band can be paused, while sensing signals are received. Future ISaC-capable RUs may support concurrent transmission and reception on the same carrier through suitable self-interference mitigation. Mechanisms to suppress interference from other carriers/sectors/sites are needed in all sensing modes. To manage QoS of communication and sensing, SQT (Sensing Quality Translation) translates the application requirements into required configuration of resources, sensing mode, etc. As the sensing impacts privacy, a concept of SePC (Sensing Policy Control) is proposed and the details on its functionality are provided. This covers aspects of sensing policies and policy enforcement as well as consent management, transparency and privacy controls. The latter shall ensure that only the minimal amount of sensing data is processed and that methods from the domain of privacy-enhancing technologies such as anonymization, and obfuscation are applied as early as possible in the overall processing chain.

**Acknowledgement**

This work has been supported by the Federal Ministry of Research, Technology and Space of the Federal Republic of Germany as part of the KOMSENS-6G project (16KISK).